\documentclass[lettersize,journal]{IEEEtran}

\usepackage{mathtools}
\usepackage{newtxmath}
\usepackage{cite}
\usepackage{algorithm}

\newcommand{\diag}[1]{\operatorname{diag}\left({#1}\right)}

\makeatletter

\makeatletter
\def\IEEElabelanchoreqn#1{\bgroup
\def\@currentlabel{\p@equation\theequation}\relax
\def\@currentHref{\@IEEEtheHrefequation}\label{#1}\relax
\Hy@raisedlink{\hyper@anchorstart{\@currentHref}}\relax
\Hy@raisedlink{\hyper@anchorend}\egroup}
\makeatother

\usepackage{dblfloatfix}
\usepackage{balance}
\usepackage[caption=false,font=footnotesize]{subfig}

\usepackage{amsmath}
\usepackage{algorithmic}
\usepackage{graphicx}
\usepackage{textcomp}
\usepackage{xcolor}
\usepackage[short,c2,nocomma]{optidef}
\usepackage{acronym}
\usepackage{textcomp}

\usepackage{siunitx}
\DeclareSIUnit{\belmilliwatt}{Bm}
\DeclareSIUnit{\dBm}{\deci\belmilliwatt}
\DeclareSIQualifier{\isotropic}{i}
\DeclareSIQualifier{\carrier}{c}

\makeatother

\usepackage{listings}
\def\BibTeX{{\rm B\kern-.05em{\sc i\kern-.025em b}\kern-.08em
		T\kern-.1667em\lower.7ex\hbox{E}\kern-.125emX}}
\begin{document}	
	\title{\LARGE Multicarrier Rate-Splitting Multiple Access: Superiority of OFDM-RSMA over OFDMA and OFDM-NOMA}
        
	\author{Mehmet Mert \c{S}ahin\IEEEmembership{, Graduate Student Member IEEE}, Onur Dizdar\IEEEmembership{, Member, IEEE}, Bruno Clerckx\IEEEmembership{, Fellow, IEEE}, Huseyin Arslan\IEEEmembership{, Fellow, IEEE}

    \thanks{
    M.M. \c{S}ahin was with the Department of Electrical Engineering, University of South Florida, Tampa, FL, 33620. He is now with Standards and Mobility Innovation Laboratory, Samsung Research America, Plano, TX, 75023, USA (e-mail: m.sahin@samsung.com). 

    O. Dizdar was with the Imperial College of London, SW7 2AZ London, U.K. He is now with VIAVI Solutions UK Ltd., SG1 2AN Stevenage, U.K. (e-mail: onur.dizdar@viavisolutions.com).
    
    B. Clerckx is with the Department of Electrical and Electronic Engineering, Imperial College London, London, SW7 2AZ, U.K. and with Silicon Austria Labs (SAL), Graz A-8010, Austria (e-mail: b.clerckx@imperial.ac.uk; bruno.clerckx@silicon-austria.com)

    H. Arslan is with the Department of Electrical and Electronics Engineering, Istanbul Medipol University, Beykoz, 34810 Istanbul, Turkey (e-mail: huseyinarslan@medipol.edu.tr).
    } 
    }
    
\maketitle
\thispagestyle{plain}
\pagestyle{plain}

\begin{abstract}
Rate-splitting multiple access (RSMA) is a multiple access technique generalizing conventional techniques, such as, space-division multiple access (SDMA),  non-orthogonal multiple
access (NOMA), and physical layer multi-casting, which aims to address   multi-user interference (MUI) in  multiple-input multiple-
output (MIMO) systems. In this study, we leverage the interference management capabilities of RSMA to tackle the issue of inter-carrier interference (ICI) in orthogonal frequency division multiplexing (OFDM) waveform. We formulate a problem to find the optimal subcarrier and power allocation for downlink transmission in a two-user system using RSMA and OFDM and propose a weighted minimum mean-square error (WMMSE)-based algorithm to obtain a solution. The sum-rate performance of the proposed OFDM-RSMA scheme is compared with that of conventional OFDM and OFDM-NOMA by numerical results. It is shown that the proposed OFDM-RSMA outperforms OFDM-NOMA and OFDMA under ICI in diverse propagation channel conditions owing to its flexible structure and robust interference management capabilities.        
\end{abstract}


\section{Introduction} \label{sec:Introduction}

Orthogonal frequency division multiplexing (OFDM) waveform has been widely studied and deployed in wireless communication standards such as 4G-LTE, 5G-NR and Wi-Fi due to its low-complexity implementation and robustness against frequency selective channel models \cite{ankarali2017waveform}. However, 'sinc' shaped subcarriers of OFDM makes it vulnerable to sources of ICI such as Doppler spread, phase noise, mismatch in local oscillators of receiver and transmitter ends, etc. In fact, ICI destroys the orthogonality of subcarriers in OFDM and causes saturation in the data rate and error floor region in the bit-error-rate (BER) analysis even though the overall transmit power of the system increases \cite{tiejun2006OFDMuDopp}. This issue needs to be addressed in order to meet demanding requirements of next generation communication standards.

Rate-splitting multiple access (RSMA) has been shown to achieve the largest upper bound for the achievable rate in the interference channel which makes it promising to address the ICI related problems of the OFDM waveform \cite{rsmaSurvey_2022}. Moreover, RSMA has been shown to encapsulate and surpass the performance of space-division multiple access (SDMA), non-orthogonal multiple access (NOMA), orthogonal multiple access (OMA), multicasting in multiple antenna networks in terms of  spectral and energy efficiency, latency, and resilience to mixed-critical quality of service  \cite{mao_clerckx_li_2018}. 

Analysis of RSMA in multi-antenna multicarrier systems has been studied by several papers  \cite{Li2020_rsmaMultiCarrier,Chen2021_rsmaMultiCarrier,dizdar2022ComJamRS}. A three step resource allocation scheme is proposed in \cite{Li2020_rsmaMultiCarrier} where power allocation on a single subcarrier, matching between user and subcarrier and power allocation among different subcarriers are solved in steps to maximize the sum-rate. In \cite{Chen2021_rsmaMultiCarrier}, RSMA is studied in the overloaded multicarrier multi-group multicast downlink scenario by formulating a joint max-min fairness and sum-rate problem. RSMA for joint communications and jamming with a multi-carrier waveform in multiple-input single-output (MISO) broadcast channel (BC) is studied in \cite{dizdar2022ComJamRS}, where optimal precoder and power allocation is investigated for simultaneous communications and jamming in cognitive radio networks. 

The abovementioned works consider RSMA to address the problems of various systems employing multi-carrier waveforms, %but not the problems of the multi-carrier waveform itself when it interacts with the wireless channel 
however, the problems of the waveform itself, such as ICI due to Doppler under mobility, are not addressed.
%the problems of ICI under high mobility and constraint on decoding order to be done at subcarrier level impeding efficiency and feasibility of multicarrier waveforms are not addressed. 
In this study, we employ RSMA to solve the challenging problems that OFDM waveform faces under practical channel conditions for the first time. Owing to the flexibility granted by the use of message-splitting and successive interference cancellation (SIC), we show that OFDM-RSMA outperforms OFDMA and OFDM-NOMA in terms of sum-rate under doubly-selective channels.

\textit{Notation:} Lower-case bold face variables indicate vectors, and upper-case bold face variables indicate matrices, $\diag{\mathbf{M}}$ returns the elements on the main diagonal of matrix $\mathbf{M}$ in a vector, $\mathcal{C}\mathcal{N}(\mu,\sigma^2)$ represents complex Gaussian random vectors with mean $\mu$ and variance $\sigma^2$. The $\ell_2$-norm of a vector and Frobenius norm of a matrix are denoted as $\Vert \cdot \Vert$ and $\Vert A \Vert_F$, respectively. $\mathbf{A} \odot \mathbf{B}$ corresponds to Hadamard multiplication of matrices $\mathbf{A}$ and $\mathbf{B}$; $\mathbf{A}^{o2}$ denote Hadamard power of matrix $\mathbf{A}$ by two. The notation $m_{ij}$ is the value located in $i$th row and $j$th column of matrix $\mathbf{M}$ and $\mathbf{e}_i$ denotes the $i$th standard unit basis vector of for $\mathbb{R}^N$.

\section{System Model} \label{sec:SystemModel}
\begin{figure}[t]
    \centering
    \subfloat[]{
    \includegraphics[width=1\linewidth]{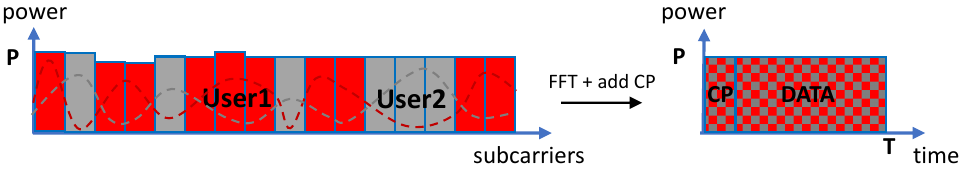}
    \label{fig:OFDMAschemes}} \hfill
    \subfloat[]{
    \includegraphics[width=1\linewidth]{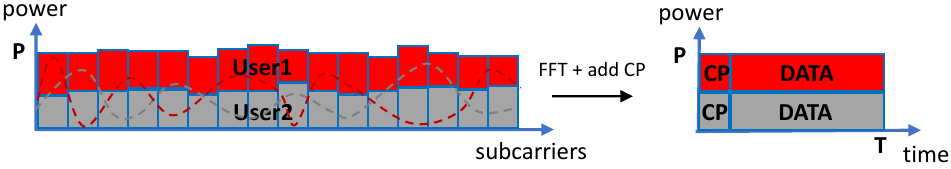}
    \label{fig:NOMAschemes}}\hfill
    \subfloat[]{
    \includegraphics[width=1\linewidth]{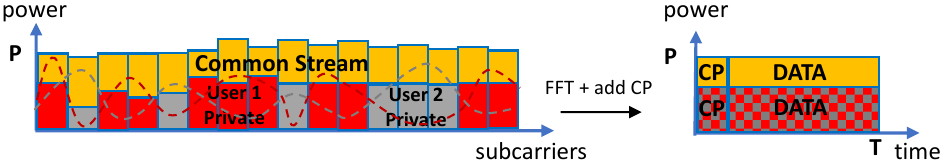}\label{fig:RSMAschemes}}
    \caption{OFDM based multiple accessing with two users showing channel power of users (dashed line) and power allocation (solid box), (a) conventional OFDMA, (b) OFDM-NOMA, (c) OFDM-RSMA.}
    \label{fig:MAschemes} 
\end{figure}

We consider a system model, where a transmitter with a single antenna serves $K$ single-antenna users. The transmitter uses OFDM to serve users in the same time slot. 
As shown in Fig. \ref{fig:MAschemes} for a two-user case, three different multiple-accessing schemes are considered for transmission {\sl i.e.,} OFDMA, OFDM-NOMA and OFDM-RSMA. 

\subsection{OFDM Transmission for $K$ users}
Let $\mathbf{x}_k$ denote the time domain signal of the $k$-th user expressed as  
\begin{IEEEeqnarray}{rCl}
    \mathbf{x}_k &=& \mathbf{A} \mathbf{F}^H \diag{\mathbf{p}_k}\mathbf{d}_k, \quad k \in \mathcal{K} = \{ 1,2,\ldots,K\}, 
    \label{eq:OFDMsignal}
\end{IEEEeqnarray}
where $\mathbf{F} \in \mathbb{C}^{N \times N}$ is the $N$-point fast Fourier transform (FFT) matrix, $\mathbf{d}_k$ is the data symbols carrying information of $k$-th user, and $\mathbf{p}_k \in \mathbb{C}^{N \times 1}$ is the precoding vector that captures the power allocated to OFDM subcarriers. The cyclic prefix (CP)-addition matrix $\mathbf{A} \in \mathbb{N}^{(N+C) \times N}$, and the CP-removal matrix $\mathbf{B} \in \mathbb{N}^{N \times (N+C)}$ can be expressed as follows:
\begin{IEEEeqnarray}{rCl} 
	\mathbf{A} &=& \begin{bmatrix}
	\mathbf{0}_{C \times (N-C)} & \mathbf{I}_{C} \\
	\multicolumn{2}{c}{\mathbf{I}_N} 
	\end{bmatrix}, \;\;\; \mathbf{B} = \begin{bmatrix}
	\mathbf{0}_{N \times C} & \mathbf{I}_N 
	\end{bmatrix}.
	\IEEEyesnumber
	\label{eq:cpAdditionMatrixGeneral} 
	\IEEEeqnarraynumspace
\end{IEEEeqnarray}
The time-domain signal $\mathbf{y}_k \in \mathbb{C}^{(N + C) \times 1}$ received through the $k$-th user channel   $\mathbf{H}_k$ can be written as 
\begin{IEEEeqnarray}{rCl} 
    \mathbf{y}_{k} &=& \mathbf{H}_{k} \sum_{k=1}^K \mathbf{x}_{k} + \mathbf{n}_k, \label{eq:transmittedSignalAfterChannel}
\end{IEEEeqnarray}
where the vector $\mathbf{n}_k \in \mathbb{C}^{(N + C) \times 1}$ is the AWGN with $n_{k,i} \sim \mathcal{CN}(0,\sigma^2)$ where $n_{k,i} \in \mathbf{n}_k$ and $\sigma^2$ is the power of AWGN.
\subsection{Channel Model}

The linear time-varying (LTV) channel model includes multipath propagation and Doppler effect leading to time and frequency shifts on the transmitted signal. The channel model includes complex channel gain, Doppler shift and delay for every path. Therefore, the LTV channel in the time-delay domain, $c(t,\tau)$, can be modeled as follows \cite{hlawatsch_matz_2011}:
\begin{IEEEeqnarray}{rCl}
  c(t,\tau)  &=& \sum_{l=1}^L \alpha_l e^{j2\pi\nu_l t}  \delta(\tau-\tau_l),
\label{eq:channelModel}  
\end{IEEEeqnarray}
where $\alpha_l$, $\tau_l$, and $\nu_l$ denote the complex attenuation factor, time delay, and Doppler frequency shift associated with the $l^{\text{th}}$ discrete propagation path where $l \in \left \lbrace 1,2,\ldots, L \right \rbrace$. Let $N$ and $C$ be the total subcarrier number with the set of $\mathcal{N} = \{ 1, 2, \ldots, N\}$ and CP length of the OFDM waveform, respectively. It is assumed that CP length is larger than the maximum delay spread to ensure nter-symbol interference
(ISI) free transmission. The relation of (\ref{eq:channelModel}) with the $k$-th user's time domain channel matrix $\mathbf{H}_k \in \mathbb{C}^{(N + C) \times (N + C)}$ can be represented as follows: 
\begin{IEEEeqnarray}{rCl}
\mathbf{H}_k  &=& \sum_{l=1}^L \alpha_l \mathbf{\Pi}^{n_{\tau_l}}  \mathbf{\Delta}(\nu_l),
\label{eq:channelModelMatrix} 
\end{IEEEeqnarray} 
where the delay matrix $\mathbf{\Pi}^{n_{\tau_l}} \in \mathbb{C}^{(N+C) \times (N+C)}$ is the forward cyclic shifted permutation matrix according to the delay of the $l$th path. The Doppler shift matrix for the $l$th path, $\mathbf{\Delta}(\nu_l) \in \mathbb{C}^{(N+C) \times (N+C)}$, is defined as $ \mathbf{\Delta}(\nu_l) = \diag{\left[e^{\frac{j2\pi\nu_l}{F_s}}, e^{\frac{j2\pi\nu_l 2}{F_s}}, \cdots, e^{\frac{j2\pi\nu_l (N+C)}{F_s}}\right]}$, where $F_s$ is the sampling frequency in the system model.
\begin{figure*}[t]
    \centering
    \includegraphics[width=1\linewidth]{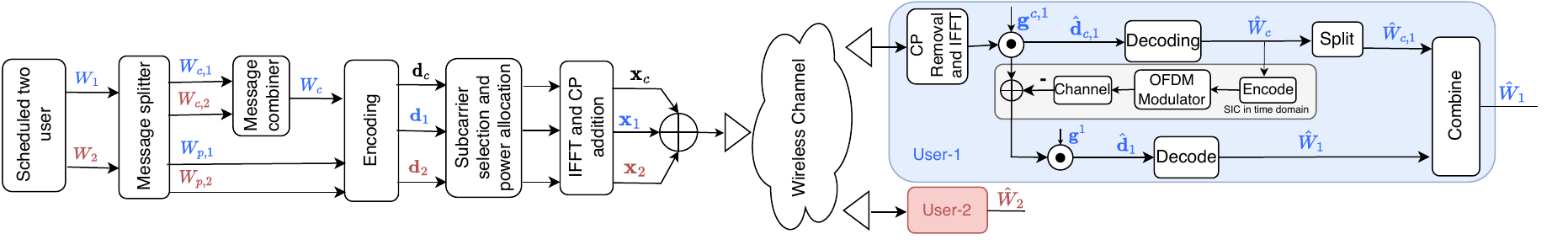}
    \caption{Proposed OFDM-RSMA scheme.} 
    \label{fig:ofdmRSMAtransceiver}
\end{figure*}

\section{Proposed OFDM-RSMA Scheme}

In this section, we describe the proposed OFDM-RSMA scheme and formulate a sum-rate maximization problem to obtain the optimal common rate, subcarrier, and power allocation for the proposed scheme. Fig. \ref{fig:ofdmRSMAtransceiver} demonstrates proposed scheme for a two-user scenario. At the transmitter, the message intended for user-$k$, $W_k$, is split into common and private parts, which are denoted as $W_{c,k}$ and $W_{p,k}$, $\forall k \in \mathcal{K}$. The common parts of messages for all users are combined into a single common message $W_{c}$. The common and private messages are independently encoded into the common and private streams, $\mathbf{d}_c$ and $\mathbf{d}_{k}, \forall k \in \mathcal{K}$, respectively. The encoded symbols $\mathbf{d}_c$ and $\mathbf{d}_{k}$ are chosen from a Gaussian alphabet for theoretical analysis. We assume that the streams have unit power, {\sl i.e.,} $\mathbb{E} \{ \tilde{\mathbf{d}}\tilde{\mathbf{d}}^H\} = \mathbf{I}$, where $\tilde{\mathbf{d}} = [\mathbf{d}_c^T, \mathbf{d}_1^T, \ldots, \mathbf{d}_K^T]^T $. 

For a $K$-user OFDM-RSMA system, the transmitted common stream $\mathbf{x}_c \in \mathbb{C}^{(N + C)\times 1}$, is expressed as follows:
\begin{IEEEeqnarray}{rCl} 
    \mathbf{x}_{c} &=& \mathbf{A}  \mathbf{F}^H \diag{\mathbf{p}_c} \mathbf{d}_c,  \label{eq:commonStream}
\end{IEEEeqnarray}
where $\mathbf{p}_{c} \in \mathbb{C}^{N \times 1}$ is the power allocation precoding vector for the common stream over the OFDM subcarriers. The private stream for user-$k$, $\mathbf{x}_k \in \mathbb{C}^{(N + C)\times 1}$, is expressed as in (\ref{eq:OFDMsignal}).
The matrix $\mathbf{P} = [\mathbf{p}_1,\ldots,\mathbf{p}_K]$ is defined as the collection of all precoding vectors of private streams, $\mathbf{p}_k, \forall k \in \mathcal{K}$. Vectors $\bar{\mathbf{p}}_{k,n}$ and $\bar{\mathbf{p}}_{c,n}$ denote that $n$-th subcarrier is forced not to carry any energy for private and common streams, i.e., $\bar{\mathbf{p}}_{k,n} = \left[ p_{k,1}, \ldots, p_{k,n-1}, 0, p_{k,n+1}, \ldots, p_{k,N} \right]$ and $\bar{\mathbf{p}}_{c,n} = \left[ p_{c,1}, \ldots, p_{c,n-1}, 0, p_{c,n+1}, \ldots, p_{c,N} \right]$. Accordingly, the time-domain received signal (\ref{eq:transmittedSignalAfterChannel}) can be re-written as follows:
\begin{IEEEeqnarray}{rCl} 
	\mathbf{y}_{k} &=& \mathbf{H}_{k} \left( \mathbf{x}_{c} + \sum_{k=1}^K \mathbf{x}_{k} \right) + \mathbf{n}_k.	\label{eq:transmittedAfterChannel}
\end{IEEEeqnarray}
At the receiver side, we first process the common stream. CP removal matrix and FFT operation are applied to the superimposed signal $\mathbf{y}_k$ to convert it into frequency domain for one-tap equalization and demodulation of the common stream. The received frequency domain signal at user-$k$, $\mathbf{r}_{c,k} = \mathbf{F} \mathbf{B}\mathbf{y}_{k}$, is expressed as follows: 
\begin{IEEEeqnarray}{rCl} 
    \mathbf{r}_{c,k} &=& \mathbf{F} \mathbf{B} \mathbf{H}_k \mathbf{A} \mathbf{F}^H \Bigg(\diag{p_{c,n} \mathbf{e}_n} \mathbf{d}_c + \diag{\bar{\mathbf{p}}_{c,n}} \mathbf{d}_c \nonumber \\ && +  \sum^K_{u=1}\diag{\mathbf{p}_u} \mathbf{d}_u \Bigg), \quad \forall n \in \mathcal{N}.	\label{eq:commonSequence_received} \IEEEeqnarraynumspace
\end{IEEEeqnarray}
The average received power at the $n$-th subcarrier of $\mathbf{r}_{c,k}$, $T_{c,k,n} \triangleq \mathsf{E} \left\{ \vert {(\mathbf{r}_{c,k})}_n \vert^2 \right\}$, is written as:
\begin{IEEEeqnarray}{rCl}
    T_{c,k,n} & = &  \left \vert s^{c,k,n}_{n,n} \right \vert^2 + \underbrace{ \sum_{j=1}^{N} \left \vert \bar{s}^{c,k,n}_{n,j} \right \vert^2 + \sum_{u = 1}^K \sum_{j=1}^{N} \left \vert s^u_{n,j} \right \vert^2 + \sigma^2}_{I_{c,k,n}}, \IEEEeqnarraynumspace \label{eq:power_common} 
\end{IEEEeqnarray}
with  
\begin{IEEEeqnarray}{rCl} 	
    \mathbf{S}^{c,k,n} &=& \mathbf{F}\mathbf{B} \mathbf{H}_k \mathbf{A} \mathbf{F}^H \diag{p_{c,n}\mathbf{e}_n}, \; \forall k \in \mathcal{K}, \; \forall n \in \mathcal{N},   \IEEEyesnumber \label{eq:signalMatrices} \IEEEyessubnumber \label{eq:intendedCommonSignal} \IEEEeqnarraynumspace \\ 
    \bar{\mathbf{S}}^{c,k,n} &=& \mathbf{F}\mathbf{B} \mathbf{H}_k \mathbf{A} \mathbf{F}^H \diag{\bar{\mathbf{p}}_{c,n}}, \; \forall k \in \mathcal{K}, \; \forall n \in \mathcal{N},   \IEEEyessubnumber \label{eq:ICISignal} \IEEEeqnarraynumspace \\
    \mathbf{S}^{u} &=& \mathbf{F}\mathbf{B} \mathbf{H}_k \mathbf{A} \mathbf{F}^H \diag{\mathbf{p}_{u}}, \quad \forall u \in \mathcal{K}, \IEEEyessubnumber \label{eq:MUISignal} \IEEEeqnarraynumspace
\end{IEEEeqnarray}
where the matrix in (\ref{eq:intendedCommonSignal}) denotes the energy on all OFDM subcarriers resulting from the transmission on $n$-th subcarrier of the common stream after passing through the LTV channel. The ICI on the $n$-th subcarrier caused by the energy leakage from the common stream on all subcarriers except the $n$-th subcarrier is written in (\ref{eq:ICISignal}). Lastly, interference due to all private streams is expressed in (\ref{eq:MUISignal}). After the common stream is demodulated at user-$k$, it is reconstructed and subtracted from the received signal. Then, FFT matrix succeeding CP removal matrix is applied to the remaining signal in order to demodulate the intended private stream for user-$k$ as follows:
\begin{IEEEeqnarray}{rCl} 
    \mathbf{r}_{k} &=& \mathbf{F} \mathbf{B}\left(\mathbf{y}_k - \mathbf{H}_k \mathbf{A} \mathbf{F}^H \diag{\mathbf{p}_c} \mathbf{d}_{c}\right) \nonumber \\ &=& \mathbf{F} \mathbf{B} \left( \mathbf{H}_k \sum^K_{k=1} \mathbf{x}_k + \mathbf{n}_k\right). \IEEEeqnarraynumspace	\label{eq:privateSequence_received}
\end{IEEEeqnarray}
The average received power at the $n$-th subcarrier of $\mathbf{r}_{k}$, $T_{k,n} \triangleq \mathsf{E} \left\{ \left\vert {(\mathbf{r}_{k})}_n \right \vert^2 \right\}$, can be written as:
\begin{IEEEeqnarray}{rCl} 
    T_{k,n} & = & \left \vert v^{k,n}_{n,n} \right \vert^2 + \underbrace{ \sum_{j=1}^{N} \left \vert \bar{v}^{k,n}_{n,j} \right \vert^2 + \sum_{\substack{i=1 \\ i \neq k}}^{K} \sum_{j=1}^{N} \left \vert w^i_{n,j}  \right \vert^2 + \sigma^2}_{I_{k,n}}, \IEEEeqnarraynumspace \label{eq:power_private} 
\end{IEEEeqnarray}	
with
\begin{IEEEeqnarray}{rCl} 	
    \mathbf{V}^{k,n} &=& \mathbf{F} \mathbf{B} \mathbf{H}_k \mathbf{A} \mathbf{F}^H \diag{p_{k,n}\mathbf{e}_n}, \; \forall k \in \mathcal{K}, \; \forall n \in \mathcal{N}, \IEEEyesnumber \label{eq:powersPrivate} \IEEEyessubnumber \label{eq:intendedPrivate} \IEEEeqnarraynumspace \\
    \bar{\mathbf{V}}^{k,n} &=& \mathbf{F} \mathbf{B} \mathbf{H}_k \mathbf{A} \mathbf{F}^H \diag{\bar{\mathbf{p}}_{k,n}}, \; \forall k \in \mathcal{K}, \; \forall n \in \mathcal{N}, \IEEEyessubnumber \label{eq:ICIprivate} \\
    \mathbf{W}^{i} &=& \mathbf{F} \mathbf{B} \mathbf{H}_k \mathbf{A} \mathbf{F}^H \diag{\mathbf{p}_i}, \; \forall i \in {\mathcal{K} \backslash k}, \IEEEyessubnumber \label{eq:MUIprivate} \IEEEeqnarraynumspace
\end{IEEEeqnarray}
where (\ref{eq:intendedPrivate}) denotes the energy of the $n$-th subcarrier of the private stream of user-$k$  spread over all subcarriers due to channel effects. The expressions (\ref{eq:ICIprivate}) and (\ref{eq:MUIprivate}) denote the ICI and multi-user interference (MUI) due to the private stream of user-$k$ and other users' private streams, respectively. By using (\ref{eq:power_common}) and (\ref{eq:power_private}), signal-to-interference-plus-noise ratios
(SINRs) of the common and private streams for a given channel state can be stated as follows: 
\begin{IEEEeqnarray}{rCl}
   \gamma_{c,k,n} &\triangleq& \left \vert s^{c,k,n}_{n,n} \right \vert^2 I^{-1}_{c,k,n} \quad \text{and} \quad \gamma_{k,n} \triangleq \left \vert v^{k,n}_{n,n} \right \vert^2  I^{-1}_{k,n}. \label{eq:SINRs}
\end{IEEEeqnarray}
The achievable rates for common stream and private streams corresponding to user-$k$ in the corresponding subcarriers can be written as follows:
\begin{IEEEeqnarray}{rCl}
   R_{c,k,n} &=& \log_2(1+\gamma_{c,k,n}) \quad \text{and} \quad R_{k,n} = \log_2(1+\gamma_{k,n}), \IEEEeqnarraynumspace \label{eq:achievableRates}
\end{IEEEeqnarray}
The requirement that common stream should be decodable by all users necessitates a constraint on the rate of common stream at the $n$th subcarrier as follows:
\begin{IEEEeqnarray}{rCl}
    R_{c,n} = \min_{k \in \mathcal{K}} R_{c,k,n}
\end{IEEEeqnarray}
% and the achievable rates for an OFDM symbol can be written as follows: 
% \begin{IEEEeqnarray}{rCl}
%    R_{c,k} &=& \sum_{n=1}^{N} R_{c,k,n}  \; \text{bit/s/Hz} \; \text{and} \; R_{k} = \sum_{n=1}^{N} R_{k,n} \; \text{bit/s/Hz}.  \label{eq:achievableRatesOFDM} \IEEEeqnarraynumspace
% \end{IEEEeqnarray}
The sum-rate maximization problem for the proposed OFDM-RSMA can be formulated as follows:
\begin{maxi!}<b>
{\mathbf{P},\mathbf{p}_c,\mathbf{c}}{ \sum^K_{k=1} \sum_{n=1}^{N} (C_{k,n}+R_{k,n}) \label{eq:optProblem1}}
{\label{eq:optProblem}}{}
\addConstraint{\sum_{n=1}^{N} R_{c,k,n}}{\geq \sum_{n=1}^{N} C_{k,n}, \quad \forall k \in \mathcal{K}, \label{eq:optProblem2}}
\addConstraint{\sum_{n=1}^N (R_{k,n} + C_{k,n})}{\geq R_k^{\text{min}}, \label{eq:optProblem3}}
\addConstraint{\Vert \mathbf{P} \Vert_F^2 + \Vert \mathbf{p}_c \Vert^2}{\leq P_t, \label{eq:optProblem4}}
\end{maxi!}
where the common rate at the $n$th subcarrier $R_{c,n}$ is shared among users such that $C_{k,n}$ represents the $k$-th user's portion of the common rate, with $R_{c,n} = \sum_{k=1}^K C_{k,n}$, and the vector $\mathbf{c} = [C_{1,1},\ldots,C_{K,1},\ldots,C_{1,N},\ldots,C_{K,N}]$ is defined as the collection of $C_{k,n}$. The first constraint, (\ref{eq:optProblem2}), ensures that each user can decode the common stream. $R_k^{\text{min}}$ defines the minimum data rate constraint for the $k$-th user used in (\ref{eq:optProblem3}). The last constraint, (\ref{eq:optProblem4}), defines the total power limitation of the system. 
% The weighted minimum mean-square error (WMMSE) based precoding optimization framework studied in \cite{joudeh_2016_SRmax_RSMA} is adopted to solve the optimization problem (\ref{eq:optProblem}). 
% \textcolor{olive}{The weighted minimum mean-square error (WMMSE) based precoding optimization framework utilizing alternating optimization is adopted to solve the optimization problem \cite{joudeh_2016_SRmax_RSMA}. In each iteration equalizers and weights over all subcarriers are updated individually using decoupling approach.}   

Let $\hat{\mathbf{d}}_{c,k} = \mathbf{g}^{c,k} \odot \mathbf{F} \mathbf{B} \mathbf{y}_k$ be $k$th user's estimate of $\mathbf{d}_{c}$, where $\mathbf{g}^{c,k} \in \mathbb{C}^{N\times 1}$ is the one tap-equalizer vector for. After successfully removing the common stream, the estimate of ${\mathbf{d}}_{k}$ can be obtained as follows:
\begin{IEEEeqnarray}{rCl}
    \hat{\mathbf{d}}_{k} &=& \mathbf{g}^{k} \odot \mathbf{F} \mathbf{B} \left(\mathbf{y}_k - \mathbf{H}_k \mathbf{A}  \mathbf{F}^H \diag{\mathbf{p}_{c}} \hat{\mathbf{d}}_{c,k}\right), \IEEEeqnarraynumspace 
\end{IEEEeqnarray}
where $\mathbf{g}^{k}$ is the corresponding one-tap equalizer vector for the $k$th user. The common and private mean-squared-errors
(MSEs) for $n$th subcarrier in the common and private stream are defined as $\varepsilon_{c,k,n} \triangleq \mathsf{E}\left\{ \left \vert \hat{d}_{c,k,n} - d_{c,n} \right \vert^2 \right\}$
and $\varepsilon_{k,n} \triangleq \mathsf{E}\left\{ \left \vert \hat{d}_{k,n} - d_{k,n} \right \vert^2 \right\}$ respectively:
\begin{IEEEeqnarray}{rCl} 
    \varepsilon_{c,k,n} &=& \left \vert g^{c,k}_n \right \vert^2 T_{c,k,n} - 2\mathfrak{R}\left\{ g^{c,k}_n s^{c,k,n}_{n,n} \right\} + 1, \IEEEyesnumber \label{eq:MSE_both} \IEEEyessubnumber \label{eq:MSE_common} \IEEEeqnarraynumspace \\
    \varepsilon_{k,n} &=& \left \vert g^{k}_n \right \vert^2 T_{k,n} - 2\mathfrak{R}\left\{ g^k_n v^{k,n}_{n,n} \right\} + 1. \IEEEyessubnumber \label{eq:MSE_private}
\end{IEEEeqnarray}
Optimum minimum mean-square error (MMSE) equalizers for corresponding subcarriers can be found by solving $\frac{\partial \varepsilon_{c,k,n}}{\partial g^{c,k}_n} = 0$ and $\frac{\partial \varepsilon_{k,n}}{\partial g^k_n} = 0$ for $n \in \mathcal{N}$, 
\begin{IEEEeqnarray}{rCl}
\left({g^{c,k}_n}\right)^{\text{MMSE}} &=& T_{c,k,n}^{-1} \left(s^{c,k,n}_{n,n}\right)^H,   \IEEEyesnumber \label{eq:Optimum_equalizer} \IEEEyessubnumber \label{eq:Optimum_equalizer_1} \\ \left({g^{k}_n}\right)^{\text{MMSE}} &=& T_{k,n}^{-1} \left({v^{k,n}_{n,n}}\right)^H. \IEEEyessubnumber \label{eq:Optimum_equalizer_2}
\end{IEEEeqnarray}
After substituting optimum equalizers found in (\ref{eq:Optimum_equalizer}) into equations in (\ref{eq:MSE_both}), MMSEs can be written as follows: 
\begin{IEEEeqnarray}{rCl} 
    \varepsilon^{\text{MMSE}}_{c,k,n} = T_{c,k,n}^{-1} I_{c,k,n}, \quad \text{and}  \quad \varepsilon_{k,n}^{\text{MMSE}} = T_{k,n}^{-1} I_{k,n}. 
    \label{eq:Optimum_MMSE} \IEEEeqnarraynumspace 
\end{IEEEeqnarray}
From (\ref{eq:Optimum_MMSE}), SINRs in (\ref{eq:SINRs}) and achievable rates in (\ref{eq:achievableRates}) are expressed as follows \cite{cioffi_2008_WMMSE}: 
\begin{IEEEeqnarray}{rCl}
    \gamma_{c,k,n} &=& \frac{1-\varepsilon^{\text{MMSE}}_{c,k,n}}{\varepsilon^{\text{MMSE}}_{c,k,n}}, \quad
    \gamma_{k,n} = \frac{1-\varepsilon_{k,n}^{\text{MMSE}}}{\varepsilon_{k,n}^{\text{MMSE}}}, \label{eq:newSINRs} \IEEEeqnarraynumspace \\ 
    R_{c,k,n} &=& -\log_2(\varepsilon^{\text{MMSE}}_{c,k,n}), \quad 
    R_{k,n} = -\log_2(\varepsilon_{k,n}^{\text{MMSE}}). \IEEEeqnarraynumspace \label{eq:newAchievableRates}
\end{IEEEeqnarray}
To solve the optimization problem of (\ref{eq:optProblem1}), the weighted
MMSE (WMMSE) method \cite{cioffi_2008_WMMSE} is adapted to the OFDM-RSMA structure. Inspired from \cite{joudeh_2016_SRmax_RSMA}, the augmented WMMSE (AWMMSE)  of common and private streams can be written as follows:
\begin{IEEEeqnarray}{rCl}
    \zeta_{c,k,n} &=& u^{c,k}_n \varepsilon_{c,k,n} - \log_2\left(u^{c,k}_n\right), \IEEEyesnumber \label{eq:AWMSEs} \IEEEyessubnumber \label{eq:AWMSEsCommon} \\  
    \zeta_{k,n} &=& u^k_n \varepsilon_{k,n} - \log_2\left(u^k_n\right), \IEEEyessubnumber \label{eq:AWMSEsPrivate}
\end{IEEEeqnarray}
where $u^{c,k}_n, u^k_n > 0$ are weights associated with the $k$th user's MSEs of common and private stream in $n$th subcarrier, respectively. Substituting optimum equalizers of into (\ref{eq:AWMSEs}) yields
\begin{IEEEeqnarray}{rCl}               
\zeta_{c,k,n}\left(\left(g^{c,k}_n\right)^{\ast}\right) &=& u^{c,k}_n\varepsilon^{\text{MMSE}}_{c,k,n} - \log_2 \left(u^{c,k}_n\right), \IEEEyesnumber \label{eq:newAWMSEs} \IEEEyessubnumber \label{eq:newAWMSEs_1} \IEEEeqnarraynumspace \\ 
\zeta_{k,n}\left(\left(g^k_n\right)^{\ast}\right) &=& u^k_n \varepsilon^{\text{MMSE}}_{k,n} - \log_2\left(u^k_n\right). \IEEEyessubnumber \label{eq:newAWMSEs_2}
\end{IEEEeqnarray}
Then, taking derivative  with respect to weight factors and equating to zero, optimum MMSE weights are obtained as follows: 
\begin{IEEEeqnarray}{rCl}
    \left(u^{c,k}_n \right)^{\ast} &=& \left(u^{c,k}_n\right)^{\text{MMSE}} \triangleq  \left(\varepsilon^{\text{MMSE}}_{c,k,n}\right)^{-1}, \IEEEyesnumber \label{eq:optimumWeights} \IEEEyessubnumber \label{eq:optimumWeights_1} \\ 
    \left(u^k_n\right)^{\ast} &=& \left(u^k_n\right)^{\text{MMSE}} \triangleq  \left(\varepsilon^{\text{MMSE}}_{k,n}\right)^{-1}. \IEEEyessubnumber \label{eq:optimumWeights_2}
\end{IEEEeqnarray}
Finally, the rate-WMMSE relationship can be formulated as follows: 
\begin{IEEEeqnarray}{rCl}
    \zeta^{\text{MMSE}}_{c,k,n} = 1- R_{c,k,n}, \quad \text{and} \quad \zeta^{\text{MMSE}}_{k,n} = 1- R_{k,n}. \label{eq:rateAWMMSE} 
\end{IEEEeqnarray}
% \begin{figure}[t]
% \begin{minipage}{1\linewidth}
% \begin{algorithm}[H] 
%  	\begin{algorithmic}[1] 
%  	\STATE      
%         \textbf{Initialize:} $n \leftarrow 0$, $\mathbf{P}^{[s]}$, $\mathbf{p}_c^{[s]}$, $\text{SR}^{[s]}$
%  	\REPEAT 
%         \STATE $n \leftarrow n+1$; 
%         \STATE $\mathbf{P}^{[s-1]} \leftarrow \mathbf{P}^{[s]}$, \text{and} $\mathbf{p}_c^{[s-1]} \leftarrow \mathbf{p}_c^{[s]}$;
%  	\STATE $\mathbf{U} \leftarrow \mathbf{U}^{\text{MMSE}}(\mathbf{P}^{[s-1]},\mathbf{p}_c^{[s-1]})$;
%  	\STATE $\mathbf{G} \leftarrow \mathbf{G}^{\text{MMSE}}(\mathbf{P}^{[s-1]},\mathbf{p}_c^{[s-1]})$;
%   \STATE update $(\bar{\boldsymbol{\zeta}^c},\mathbf{p}_c, \mathbf{P})$ by solving (\ref{eq:optProblem_2}) using the updated $\mathbf{U}$, and $\mathbf{G}$
%  	\UNTIL convergence in $\text{SR}$ is reached 
%     \end{algorithmic}
% 	\caption{Alternating Optimization for OFDM-RSMA} 
% 	\label{algo:OFDMRSMA} 
% \end{algorithm}	
% \end{minipage}
% \end{figure}
The deterministic version of AWMSE minimization problem can be formulated as follows: 
\begin{mini!}<b>
{\boldsymbol{\zeta}^c,\mathbf{P}_c,\mathbf{P},\mathbf{U},\mathbf{G}}
{\sum^K_{k=1} \sum_{n=1}^{N} \zeta_{c,k,n} + \sum^K_{k=1} \sum_{n=1}^{N} \zeta_{k,n} \label{eq:optProblem_31}}
{\label{eq:optProblem_3}}{}
\addConstraint{\sum_{n=1}^{N} \zeta_{c,k,n}}{\leq \sum_{k=1}^K \sum_{n=1}^{N} X_{c,k,n} - N(K-1),}
\addConstraint{\Vert \mathbf{P} \Vert^2_F + \Vert \mathbf{P}_c \Vert_F^2}{\leq P_t,}
\addConstraint{\sum_{n=1}^{N} \zeta_{c,k,n} +  \sum_{n=1}^{N} \zeta_{k,n}}{\leq 2N - R_k^{\text{min}},}
\end{mini!}
where $X_{c,k,n} = 1-C_{c,k,n}$ and the sets $\mathbf{G}$ and $\mathbf{U}$ are the collection of optimum MMSE equalizers and weights.  
% It is shown that  when $(\boldsymbol{\zeta}^c, \mathbf{P}_c, \mathbf{P},\mathbf{U})$ are fixed, the optimal equalizer is the MMSE equalizer $\mathbf{G}$, and when $(\boldsymbol{\zeta}^c, \mathbf{P}_c, \mathbf{P},\mathbf{G})$ are fixed, the optimal weight is the MMSE weight $\mathbf{U}$. When $(\mathbf{U},\mathbf{G})$ are fixed, $(\boldsymbol{\zeta}^c,\mathbf{P}_c, \mathbf{P})$ is coupled in the optimization
% problem (\ref{eq:optProblem_3}), closed-form solution cannot be derived. 
Alternating optimization algorithm can be used to solve (\ref{eq:optProblem_3}). In $s$th iteration of the algorithm, the equalizers and weights are firstly updated using the power allocation matrix obtained in the ($s-1$)th iteration. Then, $(\boldsymbol{\zeta}^c, \mathbf{P}_c, \mathbf{P})$ can be updated by solving the problem (\ref{eq:optProblem_3}).  $(\boldsymbol{\zeta}^c, \mathbf{P}_c, \mathbf{P})$ and $(\mathbf{U}, \mathbf{G})$ are iteratively updated until the sum-rate converges. The algorithm is guaranteed to converge as sum-rate increasing in each iteration and it is bounded above for a given power constraint. 

By fixing the precoder, the equalizers and weights are updated at the $s$th iteration of the algorithm, using common and private precoders found at the $(s-1)$th iteration. The intermediate parameters which are obtained using the updated $(\mathbf{G}, \mathbf{U})$ can be listed as follows: 
\begin{IEEEeqnarray}{rCl}
   \boldsymbol{\alpha}_{c,k} & = & \mathbf{u}^{c,k} \odot \left\vert \mathbf{g}^{c,k}  \right \vert^{o 2}, \quad  \boldsymbol{\alpha}_{k}  =  \mathbf{u}^k \odot \left \vert \mathbf{g}^k \right \vert^{o 2}, \nonumber \\ \boldsymbol{\beta}_{c,k} &=&  \diag{\left(\boldsymbol{\alpha}_{c,k}\right)^{o (1/2)}}  \mathbf{F} \mathbf{B} \mathbf{H}_k \mathbf{A} \mathbf{F}^H, \nonumber
    \\ 
    \boldsymbol{\beta}_{k}  &=& \diag{\left(\boldsymbol{\alpha}_{k}\right)^{o (1/2)}}  \mathbf{F} \mathbf{B} \mathbf{H}_k \mathbf{A} \mathbf{F}^H, \nonumber
   \\ \boldsymbol{f}_{c,k} &=& \diag{\mathbf{u}^{c,k} \odot \mathbf{g}^{c,k}}  \mathbf{F} \mathbf{B} \mathbf{H}_k \mathbf{A} \mathbf{F}^H, \nonumber
    \IEEEeqnarraynumspace \\
    \boldsymbol{f}_{k} &=& \diag{\mathbf{u}^{k} \odot \mathbf{g}^k}  \mathbf{F} \mathbf{B} \mathbf{H}_k \mathbf{A} \mathbf{F}^H, \nonumber \IEEEeqnarraynumspace \\
    \boldsymbol{\upsilon}_{c,k}  & = & \log_2\left(\mathbf{u}^{c,k}\right), \quad \boldsymbol{\upsilon}_{k} = \log_2\left(\mathbf{u}^k \right). \nonumber
\end{IEEEeqnarray}
By fixing the equalizer and weights, the optimization problem can be written as follows to update the precoder:
\begin{mini!}<b>
{\substack{\boldsymbol{\zeta}^c,\\\mathbf{P}_c,\\\mathbf{P}}}{\sum_{k=1}^K \Bigg[ \sum_{n=1}^{N} \Bigg(\zeta_{c,k,n}+\sum_{\substack{i=1 \\ i \neq k}}^K \sum_{j=1}^{N} \vert\chi^i_{n,j}\vert^2  + \vert \kappa^{k,n}_{n,n} \vert^2}
{\label{eq:UpdateWeights}}{}
\breakObjective{ + \sum_{j=1}^{N} \vert \bar{\kappa}^{k,n}_{n,j}\vert^2 + \left(\boldsymbol{\alpha}_{k}\right)_n \sigma^2  - 2\mathfrak{R}\left\{ \omega^k_n \right\} + u^k_n - \upsilon^k_n \Bigg) \Bigg] \nonumber}
\addConstraint{\sum_{n=1}^{N} \Bigg( \sum_{i=1}^K  \sum_{n=1}^{N} \vert \psi^{i}_{n,n} \vert^2 +  \vert \phi^{k,n}_{n,n} \vert^2 + \sum_{j=1}^{N} \vert \bar{\phi}^{k,n}_{n,j} \vert^2}{}
\addConstraint{+\left(\boldsymbol{\alpha}_{c,k}\right)_n + \sigma^2  - 2\mathfrak{R}\left\{ \omega^{c,k}_{n}\right\} + u^{c,k}_n + v^{c,k}_n \Bigg)}{\leq \zeta_{c} \nonumber}{}
\addConstraint{\Vert \mathbf{P} \Vert^2_F + \Vert \mathbf{p}_c \Vert^2}{\leq P_t,}
\addConstraint{\sum_{n=1}^{N} \zeta_{c,k,n} +  \sum_{n=1}^{N} \zeta_{k,n}}{\leq 2N - R_k^{\text{min}}},
\end{mini!}
where $\zeta_{c} = \sum_{k=1}^K \sum_{n=1}^{N} \zeta_{c,k,n}$ and
\begin{IEEEeqnarray}{rCl} 	
    \boldsymbol{\chi}^i &=& \boldsymbol{\beta}_i \diag{\mathbf{p}_i}, \quad \boldsymbol{\psi}^{i} = \boldsymbol{\beta}_{c,i} \diag{\mathbf{p}_i}, \nonumber \\ 
    \boldsymbol{\kappa}^{k,n} &=& \boldsymbol{\beta}_k \diag{\mathbf{p}_{k,n}}, \quad \bar{\boldsymbol{\kappa}}^{k,n} = \boldsymbol{\beta}_k \diag{\bar{\mathbf{p}}_{k,n}}, \nonumber \\ \boldsymbol{\omega}^k &=& \diag{\mathbf{f}_{k}\diag{ \mathbf{p}_k}}, \quad \boldsymbol{\omega}^{c,k} =\diag{\mathbf{f}_{c,k} \diag{\mathbf{p}_{c}}}, \nonumber \\  
    \boldsymbol{\phi}^{k,n} &=& \boldsymbol{\beta}_{c,k} \diag{\mathbf{p}_{c,n}}, \quad \bar{\boldsymbol{\phi}}^{k,n} = \boldsymbol{\beta}_{c,k} \diag{\bar{\mathbf{p}}_{c,n}}. \nonumber
\end{IEEEeqnarray}

\section{OFDM-NOMA transmission}
In a $K$-user OFDM-NOMA system, $k$-th user message is decoded after $k-1$ users' signals are successively canceled. SIC order is not alternated at subcarrier level but kept fixed throughout one OFDM symbol as done in practical systems. The received frequency domain signal at the $k$-th user, $\mathbf{r}_{k}$ in (\ref{eq:privateSequence_received}), is modified as follows:
\begin{IEEEeqnarray}{rCl} 
\mathbf{r}_{k} &=& \mathbf{F} \mathbf{B}\left(\mathbf{y}_{k} - \mathbf{H}_k \mathbf{A} \mathbf{F}^H \left(\sum_{l=1}^{k-1}  \diag{\mathbf{p}_l}\mathbf{d}_l \right)\right).	\label{eq:nomaReceived}
\end{IEEEeqnarray}
The received power, $T_{k,n} \triangleq \mathsf{E} \left\{ \left\vert {(\mathbf{r}_{k})}_n \right \vert^2 \right\}$, at the $n$th subcarrier of $\mathbf{r}_{k}$ can be written as follows:
\begin{IEEEeqnarray}{rCl} 
    T_{k,n} & = & \left \vert v^{k,n}_{n,n} \right \vert^2 + \underbrace{ \sum_{j=1}^{N} \left \vert \bar{v}^{k,n}_{n,j} \right \vert^2 + \sum_{\substack{i=k+1}}^{K} \sum_{j=1}^{N} \left \vert w^i_{n,j}  \right \vert^2 + \sigma^2}_{I_{k,n}}, \IEEEeqnarraynumspace \label{eq:NOMApower} 
\end{IEEEeqnarray}	
where
\begin{IEEEeqnarray}{rCl} 	
    \mathbf{V}^{k,n} &=& \mathbf{F} \mathbf{B} \mathbf{H}_k \mathbf{A} \mathbf{F}^H \diag{p_{k,n}\mathbf{e}_n}, \; \forall k \in \mathcal{K}, \; \forall n \in \mathcal{N},  \nonumber  \IEEEeqnarraynumspace \\
    \bar{\mathbf{V}}^{k,n} &=& \mathbf{F} \mathbf{B} \mathbf{H}_k \mathbf{A} \mathbf{F}^H \diag{\bar{\mathbf{p}}_{k,n}}, \; \forall k \in \mathcal{K}, \; \forall n \in \mathcal{N}, \nonumber \IEEEeqnarraynumspace \\
    \mathbf{W}^{i} &=& \mathbf{F} \mathbf{B} \mathbf{H}_k \mathbf{A} \mathbf{F}^H \diag{\mathbf{p}_i}, \; \forall i \in \{k+1, \ldots, K\}. \nonumber \IEEEeqnarraynumspace
\end{IEEEeqnarray}
Then, the optimization problem for achievable rate maximization using OFDM-NOMA is formulated as follows: 
\begin{maxi!}<b>
{\mathbf{P}}{ \sum^K_{k=1} \sum_{n=1}^{N} R_{k,n}}
{\label{eq:NOMAoptProblem}}{}
\addConstraint{\sum_{n=1}^N R_{k,n}}{\geq R_k^{\text{min}}},
\addConstraint{\Vert \mathbf{P} \Vert_F^2}{\leq P_t},
\end{maxi!} 
where the matrix $\mathbf{P} = [\mathbf{p}_1,\ldots,\mathbf{p}_K]$ is defined as the collection of all users' precoding vector, $\mathbf{p}_k, \forall k \in \mathcal{K}$.
The formulated problem can be solved using the WMMSE-based approach similar to the technique used for the proposed OFDM-RSMA scheme.

\section{Simulation Results}

In this section, we demonstrate the performance gain of the proposed OFDM-RSMA scheme over OFDMA and OFDM-NOMA under different channel conditions including flat fading, frequency and time selectivity. For simplicity, we study the scenarios where $K=2$. For  OFDM-NOMA, it is assumed that user-$1$ is the weak user having a smaller overall channel gain than user-$2$ (strong user), so that, the signal of user-$1$ is decoded first in the SIC process \cite{maraqa2020NOMAsurvey}. Throughout the simulations, the OFDM waveform has $35$ subcarriers with a sub-carrier spacing (SCS) of \SI{60}{\kilo\hertz}.

\begin{figure}[t]
\centering\includegraphics[width=0.8\columnwidth]{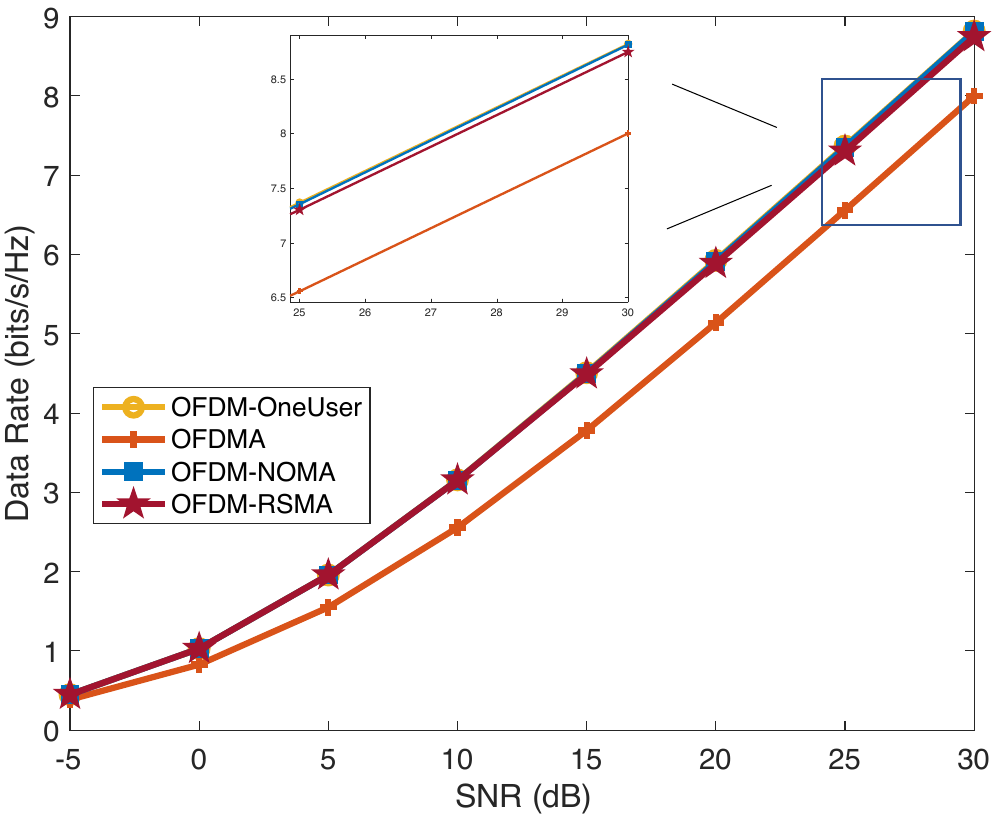}
    \caption{Performance comparison of OFDMA, OFDM-NOMA, OFDM-RSMA and one user OFDM under flat fading channel.} 
    \label{fig:flatFading}
\end{figure}

Fig. \ref{fig:flatFading} illustrates the performance of the considered multiple-accessing schemes under flat-fading channel without Doppler. Here, the sum-rate performance of the proposed OFDM-RSMA method is compared to OFDM with a single user utilizing the whole bandwidth, OFDMA with two users where the whole bandwidth is divided into two equal parts, and OFDM-NOMA. Since flat fading channel with OFDM waveform can be seen as single pipeline SISO-BC scenario, maximizing the sum-rate in OFDM-NOMA and OFDM-RSMA results in allocating power to the strongest user, and hence, performing single user OFDM \cite{clerckx_2020_MAcomparison}. The gain of OFDM-RSMA and OFDM-NOMA over OFDMA emanates from the bandwidth division among users, which makes  OFDMA a suboptimal transmission strategy from information-theoretic perspective \cite{Liu2022_NGMAnoma}.

\begin{figure}[t]
    \centering
    \includegraphics[width=0.8\columnwidth]{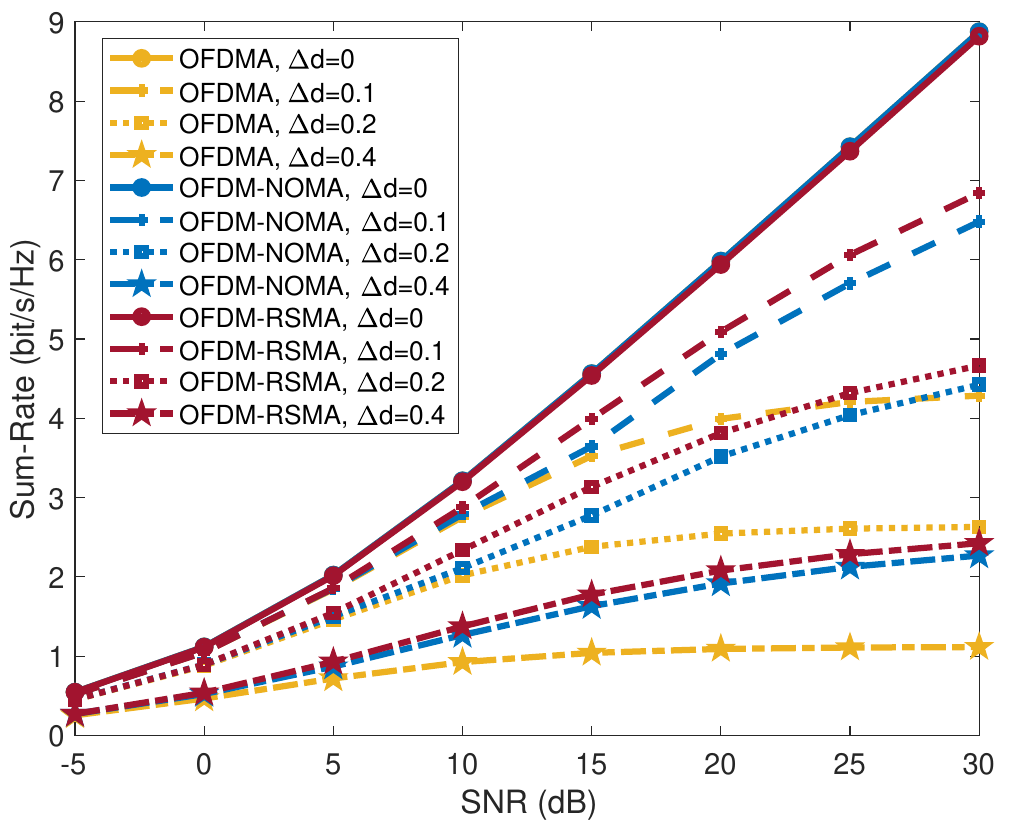} 
    \caption{Sum-rate comparison of OFDMA, OFDM-NOMA, and OFDM-RSMA under the frequency selective and doubly selective channels.} \label{fig:doublySelectiveLetter}
\end{figure}

Fig. \ref{fig:doublySelectiveLetter} demonstrates the sum-rate performance of OFDMA, OFDM-NOMA, and OFDM-RSMA under the both frequency selective and doubly dispersive channels with varying $\Delta d = \frac{f_d}{\Delta f}$, where $f_d$ is the maximum Doppler spread and $\Delta f$ is the subcarrier spacing. When the channel becomes a frequency selective channel without Doppler spread, $\Delta d=0$,
%, and the proposed OFDM-RSMA scheme provides more than 5\% sum-rate gain compared to the OFDM-NOMA. Such a gain is achieved owing to the message-splitting and the deterministic decoding order in SIC for OFDM-RSMA, enabling it to overcome the power variations over subcarriers as opposed to OFDM-NOMA, which requires switching decoding order at each subcarrier for optimal performance. Furthermore, OFDM-RSMA 
the OFDM-NOMA and proposed OFDM-RSMA schemes achieve the same performance with waterfilling based OFDMA, which is known to be capacity achieving in frequency selective channels without mobility \cite{Tse_fundamentalsOfWirelessComm_2005}. As ICI increases in doubly selective channels, the sum-rate of OFDMA drops sharply and saturates at the high SNR regime when interference becomes more dominant than the noise level. It can be seen that OFDM-RSMA and OFDM-NOMA achieve higher data rate than OFDMA in this case due to the SIC process. At high SNR regime, the proposed OFDM-RSMA achieves a higher sum-rate due to its better interference management capability compared to the OFDM-NOMA.  
%One can also conclude that the gain over OFDM-NOMA depends on the total subcarrier number in an OFDM symbol, and the maximum Delay spread of the channel.   

\section{Conclusions} \label{sec:Conclusion}

In this letter, we consider RSMA to address the problems of OFDM waveform under LTV channels. The proposed OFDM-RSMA scheme is robust against ICI stemming from time variations and outperforms OFDMA. Additionally, it is shown that inefficient use of SIC in the OFDM-NOMA scheme limits the exploitation of power variation over subcarriers, a problem which OFDM-RSMA tackles owing to its message-splitting framework. The results show that OFDM-RSMA can provide robustness against performance-limiting challenges of wireless propagation channel, such as, ISI, MUI, ICI, and inter-numerology interference (INI).   

\bibliographystyle{IEEEtran}
\bibliography{references}
	
\end{document}